# Enhanced trion emission in monolayer MoSe$_2$ by constructing a type-I van der Waals heterostructure

*Juanmei Duan [1,2,\*], Phanish Chava[1,2], Mahdi Ghorbani-Asl[1], Denise Erb[1], Liang Hu[4], Arkady V. Krasheninnikov[1,5], Harald Schneider[1], Lars Rebohle[1], Artur Erbe[1], Manfred Helm[1,2], Yu-Jia Zeng[3,\*], Shengqiang Zhou[1,\*] and Slawomir Prucnal[1,\*]*

[1]Helmholtz-Zentrum Dresden-Rossendorf, Institute of Ion Beam Physics and Materials Research, Bautzner Landstrasse 400, D-01328 Dresden, Germany

[2]Technische Universität Dresden, D-01062 Dresden, Germany

[3]College of Physics and Optoelectronic Engineering, Shenzhen University, 518060, Shenzhen, P. R. China

[4]Key Laboratory of Novel Materials for Sensor of Zhejiang Province, Institute of Advanced Magnetic Materials, Hangzhou Dianzi University, Hangzhou 310018, P. R. China

[5]Department of Applied Physics, Aalto University School of Science, Aalto FI- 00076, Finland

E-mail: juanmei.duan@hzdr.de, s.zhou@hzdr.de and s.prucnal@hzdr.de

**Abstract:** Trions, quasi-particles consisting of two electrons combined with one hole or of two holes with one electron, have recently been observed in transition metal dichalcogenides (TMDCs) and drawn increasing attention due to potential applications of these materials in light-emitting diodes, valleytronic devices as well as for being a testbed for understanding many-body phenomena. Therefore, it is important to enhance the trion emission and its stability. In this study, we construct a MoSe$_2$/FePS$_3$ van der Waals heterostructure (vdWH) with type-I band alignment, which allows for carriers injection from FePS$_3$ to MoSe$_2$. At low temperatures, the neutral exciton (X$^0$) emission in this vdWH is almost completely suppressed. The I$_{Trion}$/I$_{x0}$ intensity ratio increases from 0.44 in a single MoSe$_2$ monolayer to 20 in this heterostructure with the trion charging state changing from negative in the monolayer to positive in the heterostructure. The optical pumping with circularly polarized light shows a 14% polarization for the trion emission in MoSe$_2$/FePS$_3$. Moreover, forming such type-I vdWH also gives rise to a 20-fold enhancement of the room temperature photoluminescence from monolayer MoSe$_2$. Our results demonstrate a novel approach to convert excitons to trions in monolayer 2D TMDCs via interlayer doping effect using type-I band alignment in vdWH.

# 1. Introduction

Two-dimensional (2D) transition metal dichalcogenides (TMDCs), $MX_2$ (M=Mo, W; X=S, Se, Te) with a hexagonal crystal structure are an exciting class of materials.[1-3] Monolayer TMDs feature a direct bandgap, a large binding energy of neutral exciton ($X^0$) of up to several hundred meV because of the reduced dielectric screening, valley-selective optical coupling due to the lack of inversion symmetry and a strong spin-orbit coupling.[4] Excitons in TMDCs can often trap an electron ($X^-$) or a hole ($X^+$) to form so-called trions.[4-7] With these exotic features and tunable bandgaps from visible to near-infrared regions, TMDCs are considered as a highly desired material class for the next generation optoelectronic and valley-based electronic applications.[8]

Stacking 2D layered materials with pre-selected properties will give rise to the formation of van der Waals heterostructures (vdWHs) with atomically sharp and near-defect-free interfaces, which may exhibit novel physics and possess versatile properties.[9-11] So far, the most extensively studied TMDC vdWHs include $MoS_2$-$MoSe_2$,[12] $MoS_2$-$WS_2$,[13-14] $MoS_2$-$WSe_2$,[15-16] $MoSe_2$-$WSe_2$,[17] $MoSe_2$-$WS_2$ [18] and $WS_2$-$WSe_2$.[19] Their band alignment is found to be staggered type-II, where the conduction band minimum (CBM) and valence band maximum (VBM) are located in the wider-bandgap (WBG) and narrower-bandgap (NBG) materials, respectively. Electrons and holes can be spatially separated, which results in the formation of the interlayer excitons and consequently in quenching of the photoluminescence (PL) in their component layers.[13]

However, the type-I band alignment, which is common in conventional semiconductor heterostructures like GaAs-AlGaAs, is very rarely reported in vdWHs. The CBM and VBM of type-I band alignment are both located in the NBG layer.[20] Therefore, in the heterojunction with type-I band alignment, carriers will flow only from the WBG to the NBG material. Consequently, in the NBG layer, the carrier concentration increases and its PL intensity is enhanced.[21-23] Nevertheless, besides the use of hexagonal boron nitride (h-BN) for passivation, up to now only a few implementations of the type-I band alignment in 2D materials, like $PbI_2$/$WS_2$ [23] and $MoS_2$/$ReS_2$,[20] $MoS_2$/ZnO-QDs,[24] have been reported. Those papers focus on the charge transfer process at the interface, which leads to a PL enhancement of the NBG layer. However, trion and exciton emission behavior in type-I vdWH with TMCDs has not yet been investigated. The conversion from excitons to trions controlled by charge density can influence the exciton lifetime and leads to a large valley polarization via valley-selective optical pumping.[25] Since in straddling bands carriers flow only in one direction, by proper selection of WBG materials (*n*-type or *p*-type) the charging state of trion can be controlled. Moreover, positively charged trions, X$^+$, have a much longer dephasing time than negatively charged trions, X$^-$ and thus a smaller linewidth, which makes them attractive for quantum technology.[26]

In this work, we demonstrate the manipulation of the trion charging state and population in monolayer MoSe$_2$ by constructing a vdWH MoSe$_2$/FePS$_3$ with type-I band alignment. FePS$_3$ with an indirect bandgap is chosen as the WBG semiconductor. The type-I band alignment is supported by density-functional theory (DFT) calculation, such that MoSe$_2$ with the narrower bandgap acts as the carrier extraction layer and *p*-type FePS$_3$ is a source of holes. We have achieved an enhancement of the trion emission with the $I_T/I_{X0}$ ratio increasing from 0.44 to 20, and a 20-fold enhancement of PL emission from the MoSe$_2$ monolayer due to carrier injection from FePS$_3$. Moreover, both the exciton emission in a single MoSe$_2$ monolayer and the trion emission in MoSe$_2$/FePS$_3$ depend on the chirality of the excitation light, indicating the valley polarization selectivity. Our study provides essential insight into the underlying carrier transport mechanism and points to potential device application based on type-I band alignment vdWH.

## 2. Results and discussion

### 2.1. Raman results

**Figure 1**(a) shows the schematic configuration of MoSe$_2$/FePS$_3$ vdWH covered with a few layers of h-BN. Except protecting samples from air, the h-BN encapsulation also reduces the linewidth of the trion and exciton emission from the MoSe$_2$ monolayer.[26] The MoSe$_2$ monolayer was fully laying on the top of FePS$_3$ and the heterostructure area is around 100 μm$^2$ which is much larger than the laser spot size (about 7 μm$^2$). Therefore, all the optical signals (including micro-Raman and micro-PL) are coming from the heterojunction. For the reference, individual MoSe$_2$ monolayers and FePS$_3$ flakes covered with h-BN were prepared as well. Figure 1(b) shows the optical microscope image of MoSe$_2$/FePS$_3$ vdWH, where regions with magenta and orange outlines represent FePS$_3$ and MoSe$_2$, respectively.

Figure 1(c) and (d) show the Raman spectra of monolayer MoSe$_2$ combined with 22 nm (corresponding to 31 layers) FePS$_3$, defined as 1ML MoSe$_2$/31ML FePS$_3$, multilayer FePS$_3$, and individual MoSe$_2$ flakes with different thicknesses (1ML, 2ML and multilayer). For MoSe$_2$, the A$_{1g}$ mode is an out-of-plane vibration modes in which Se atoms in all layers oscillate in phase with reference to the corresponding central Mo atom.[27] Figure 1(d) shows the micro-Raman spectra taken from the MoSe$_2$ layer with different thicknesses. The peak positions of the A$_{1g}$ mode at 240.6 cm$^{-1}$ for 1ML MoSe$_2$, 241.7 cm$^{-1}$ for 2ML and 242.3 cm$^{-1}$ for multilayer MoSe$_2$ are in good agreement with previous reports,[27] confirming the thickness of MoSe$_2$ in our study. Besides the A$_{1g}$ phonon mode, the MoSe$_2$ flake has other Raman active phonon modes like $E_{2g}^1$ and $B_{2g}^1$ visible at about 286.9 cm$^{-1}$ and 355.7 cm$^{-1}$, as well as $E_{1g}$ at 169.7 cm$^{-1}$ for 1ML MoSe$_2$ (see Figure 1(c)). Their peak positions shift to lower wavenumbers as the number of layers increases and the highest peak intensities are observed for the MoSe$_2$ bilayer. The other peaks labeled in Figure 1(c) with a, b and c showing much broader features especially pronounced in monolayer and bilayer MoSe$_2$ are assigned to second-order Raman processes.[28] Interestingly the A$_{1g}$ phonon mode from 1ML MoSe$_2$ in vdWH has the same peak position as that from individual 1ML MoSe$_2$, but the peak intensity is significantly reduced, probably due to the softening of the phonons by phonon-carrier interaction.

The Raman spectrum from FePS$_3$ is also presented in Figure 1(c). FePS$_3$ shows six distinguishable peaks labeled with P$_1$-P$_6$. The observed phonon modes are in agreement with previously reported ones.[29] The P$_3$-P$_6$ phonon modes are assigned to the (P$_2$S$_6$)$^{4-}$ bipyramid structures and P$_1$-P$_2$ are related to the Fe atoms and are sensitive to the magnetic coupling between Fe atoms.[30] According to the temperature dependent micro-Raman spectra the evolution of the P$_1$ and P$_2$ phonon modes in FePS$_3$ is very similar both in the individual flake and in the vdWH. We found that the Néel temperature of FePS$_3$ is at around 120 K (see Figure S1), which is in good agreement with the data reported in the literature.[29]

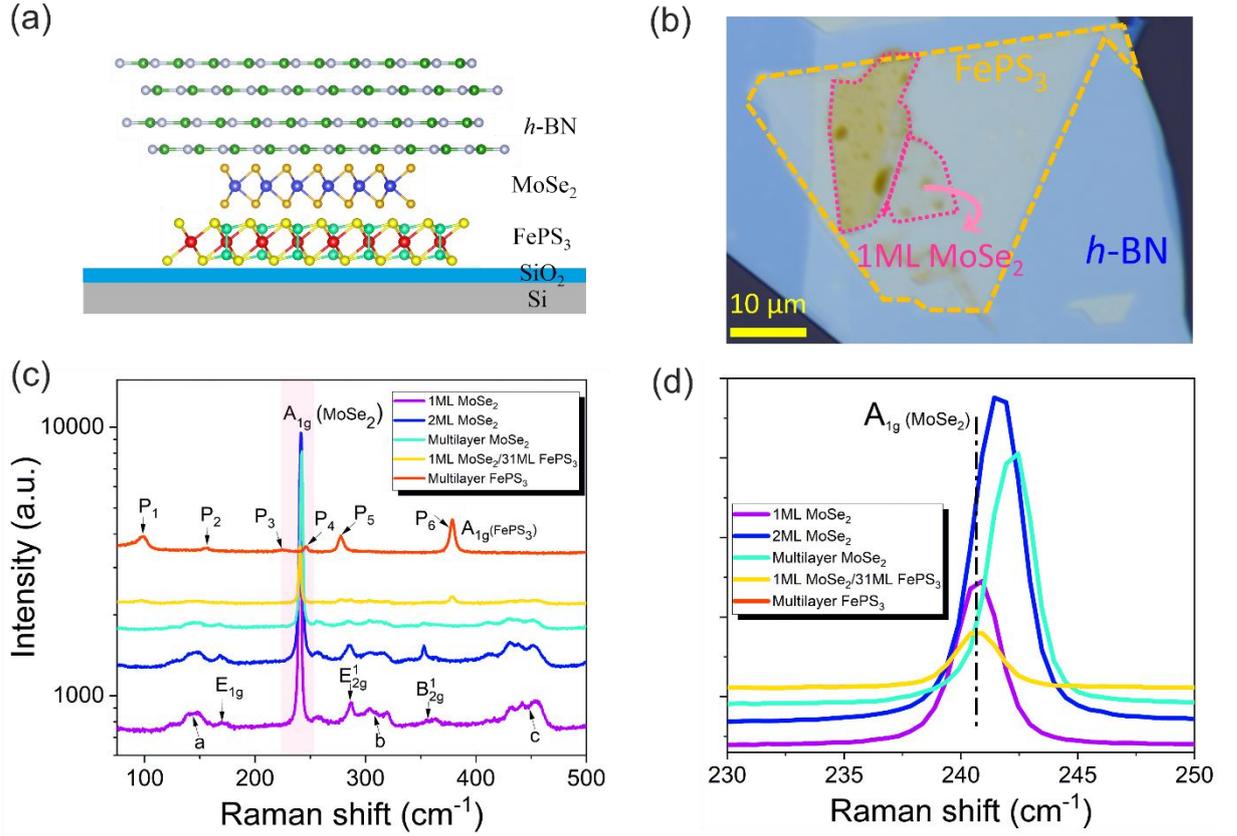

**Fig. 1. (a)** Schematic diagram of a fully *h*-BN covered MoSe$_2$/FePS$_3$ vdWH on a SiO$_2$/Si substrate. **(b)** Optical image of the 1ML MoSe$_2$ placed on the 31ML FePS$_3$ in vdWH. **(c)** Raman spectra from the 1ML MoSe$_2$, 2ML MoSe$_2$, multilayer MoSe$_2$, 1ML MoSe$_2$/31ML FePS$_3$ vdWH and multilayer FePS$_3$. **(d)** Zoomed-in Raman spectra for the marked region in panel (c) in the range of 230-250 cm$^{-1}$ under a 532 nm laser excitation. All spectra were shifted vertically for clarity.

### 2.2. Band alignment from DFT calculations

We used DFT calculations to investigate the band alignment for monolayer MoSe$_2$/FePS$_3$. For the calculations, we have considered a monolayer heterojunction 1ML MoSe$_2$/1ML FePS$_3$ and a heterojunction prepared of 3ML MoSe$_2$ on monolayer FePS$_3$, as shown in **Figure 2** (a). After structural optimization, an average interlayer distance of $d = 3.63$ Å was found between FePS$_3$ and MoSe$_2$. The binding energy between FePS$_3$ and MoSe$_2$ was calculated to be 16 meV/Å$^2$ which is close to the values in other 2D vdWH interfaces.[31] Our electronic structure calculations indicate that the MoSe$_2$ monolayer has a direct band gap of 1.33 eV at the K point (Figure S2). The FePS$_3$ monolayer shows a quasi-direct bandgap of 2.27 eV where VBM and CBM are both located at the Σ point (midpoint of Γ and K points). Although DFT band gap cannot directly be compared to the optical gap measured in the experiment, our results are in line with the experimental band gap (2.2 eV)[32] and the theoretical band gap (2.5 eV)[33] for FePS$_3$ monolayer. The slight difference of band gap values for FePS$_3$ monolayer between our calculation and the previous theoretical report [33] is due to different calculation methods. Based on the electronic structure calculations, the CBM of the MoSe$_2$ is lower than that of the FePS$_3$ while the VBM is higher than that of the FePS$_3$, forming type-I band alignment, as schematically illustrated in Figure 2 (b, c). Both heterostructures with 1ML-MoSe$_2$ and 3ML-MoSe$_2$ show direct band gaps of 1.51 eV and 1.07 eV, situated at the Γ point in Figure 2 (b, c). The projected densities of states suggest that the VBM mainly consists of Mo *d* and Se *p* orbitals, whereas the CBM is mainly composed of the Mo *d* orbitals and slight contributions from Fe *d* states. To further analyze the effect of coupling between FePS$_3$ and MoSe$_2$, the average potential profile across the heterostructure and the charge difference between the combined system and isolated parts are shown in Figure S3. Owing to the higher potential energy of MoSe$_2$, the charge is redistributed in the system with predominant accumulation at the side of MoSe$_2$ layer facing FePS$_3$.

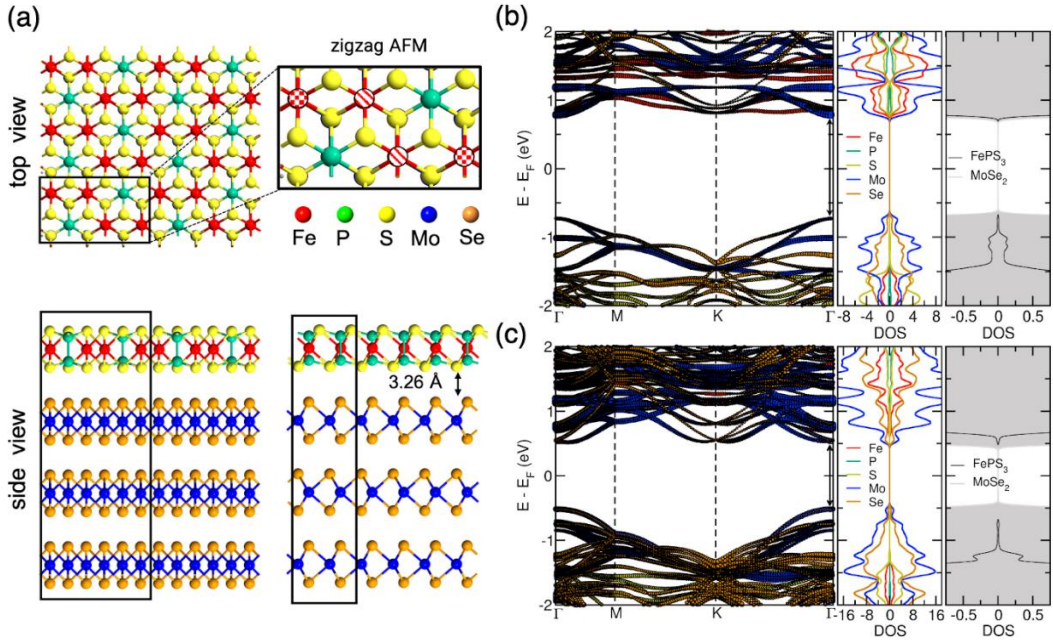

**Fig. 2. (a)** Atomic structure of FePS$_3$/MoSe$_2$, the top and side views. The rectangular unit cell of FePS$_3$ is indicated by black lines. The hatched red markers illustrate the directions of local spins (up and down, checkerboard and striped circles) of zigzag antiferromagnetic order in the unit cell. **(b)** Calculated electronic band structure and density of states (DOS) of 1ML MoSe$_2$/1ML FePS$_3$ and **(c)** 3ML MoSe$_2$/1ML FePS$_3$ by DFT.

## 2.3. Photoluminescence results

**Figure 3** (a) shows the room temperature PL (RTPL) spectra of 1ML MoSe$_2$/31ML FePS$_3$, and individual MoSe$_2$ monolayer. The RTPL spectra obtained from 2ML individual MoSe$_2$ and its heterostructure are presented in Figure 3(b). Note that the RTPL measurement was performed in air under ambient conditions. Insets show optical images of investigated samples. 1ML and 2ML $MoSe_2$ are specially chosen here since $MoSe_2$ possesses thickness-dependent bandgap features,[2, 34] i.e., a direct bandgap for monolayer $MoSe_2$ and an indirect bandgap for thicker $MoSe_2$. Similar to Raman spectra, the PL measurements can be used to determine the thickness of investigated $MoSe_2$ flakes. All the samples for PL measurement were covered with h-BN in order to prevent oxygen/moisture contaminations and obtain a high quality of optical emission spectra.[35] Since $FePS_3$ is an indirect band gap semiconductor[33] and its photogenerated carriers can quickly relax to $MoSe_2$ due to their type-I band alignment, all the PL peaks presented here originate only from $MoSe_2$ flakes. The single $MoSe_2$ monolayer flake shows the RT excitonic PL emission at 1.573 eV while the PL emission from 1ML $MoSe_2/FePS_3$ is slightly blue-shifted to 1.579 eV. More importantly, the heterostructure shows 20-fold enhancement of PL emission. Since the exciton lifetime in indirect semiconductors is generally much longer than in direct ones, the photogenerated carriers in $FePS_3$ can quickly diffuse into $MoSe_2$ and radiatively recombine there, which accounts for this giant PL enhancement in 1ML $MoSe_2$/31ML $FePS_3$. The PL emission from bilayer $MoSe_2$ on $FePS_3$ observed at 1.543 eV is enhanced by a factor of 1.3 in comparison to the individual flake. Zhang et al. have shown strong enhancement of the exciton emission from the $WS_2$ monolayer covered with $WO_3$ due to the reduction of non-radiative recombination and the charge transfer within the heterostructure .[36]

As concluded from DFT calculation, 1ML $MoSe_2/FePS_3$ shows a straddling type-I band alignment between $MoSe_2$ (NBG) and $FePS_3$ (WBG). In addition, the conductivity type of single $MoSe_2$ and $FePS_3$ flakes is confirmed by electrical measurements presented in Figure S4. $MoSe_2$ and $FePS_3$ possess *n*-type and *p*-type conductivity, respectively, which is consistent with other reports.[37, 38] After forming a heterojunction with type-I band alignment, holes will flow from $FePS_3$ to $MoSe_2$ to achieve an equilibrium state (see the middle panel in Figure 3(c)). In our case the thickness of the $MoSe_2$ is limited to one monolayer (about 0.7 nm), which can be completely converted to *p*-type when combining with 31 ML $FePS_3$. The shallowed areas in Figure 3(c) indicate the band alignment for 1ML $MoSe_2$/multilayer $FePS_3$. Under excitation, both electrons and holes generated in $FePS_3$ can diffuse to $MoSe_2$ with a result of its PL enhancement due to type-I band alignment (see the right panel in Figure 3(c)).

The smaller enhancement in 2ML $MoSe_2$/24ML $FePS_3$ is due to the fact that 2ML $MoSe_2$ is an indirect bandgap semiconductor with much lower radiative recombination rate.[34] Since the $MoSe_2$ layer is placed fully on the $FePS_3$ flake, we can exclude the possibility that surface charges in the SiO$_2$ substrate can donate carriers into MoSe$_2$, as it was previously reported.[35, 39] Therefore, in our case only FePS$_3$ can be the extra source of photogenerated carriers in MoSe$_2$. A PL emission enhancement for the narrower bandgap material in type-I van der Waals heterojunctions has been reported before, e.g. for MoTe$_2$-WSe$_2$,[21] PbI$_2$-WS$_2$,[40] WSe$_2$-black phosphorus,[41] MoS$_2$-ReS$_2$,[20] GaSe-GaTe,[22] etc. This distinct feature is substantially different from that of the widespread type-II band alignment for TMDCs heterostructures, like MoS$_2$-WSe$_2$,[15] MoS$_2$-MoSe$_2$,[12] MoS$_2$-WS$_2$ and WSe$_2$-MoS$_2$,[13, 42] where the PL emissions of two constituent layers are both quenched along with the appearance of an additional interlayer exciton peak with low energy.[17, 43]

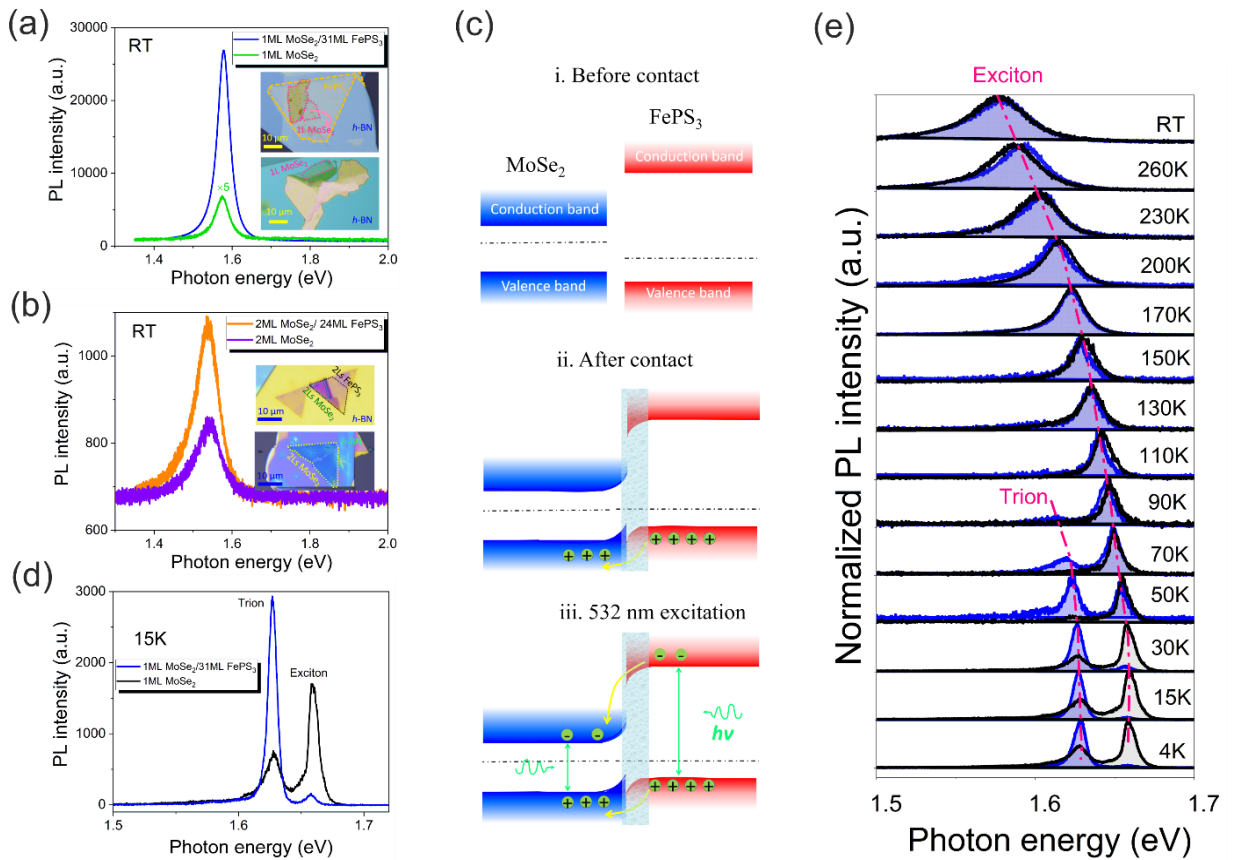

**Fig. 3.** (a) RTPL spectra of 1ML MoSe$_2$/31ML FePS$_3$ and 1ML MoSe$_2$-single layer. (b) RTPL spectra of 2ML MoSe$_2$/24ML FePS$_3$ and 2ML MoSe$_2$-single layer. Insets are optical images of samples. (c) Band alignment of MoSe$_2$/ FePS$_3$: before contact of constituent layers, after contact of constituent layers and under laser excitation. The shadowed areas visualize the band alignment for 1ML MoSe$_2$/ multilayer FePS$_3$. (d) PL spectra of 1ML MoSe$_2$/31ML FePS$_3$ and single 1ML MoSe$_2$ at 15 K. (e) Normalized TDPL spectra for 1ML MoSe$_2$/31ML FePS$_3$ (blue-shaded areas) and 1ML MoSe$_2$ (black-shaded areas).

To shine more light on the interaction and charge transfer between FePS$_3$ and MoSe$_2$, temperature-dependent PL (TDPL) measurements were performed. In this measurement, we focus on the sample 1ML MoSe$_2$/31ML FePS$_3$. As the reference sample, an individual MoSe$_2$ monolayer was also investigated. Figure 3 (d) shows the PL spectra of 1ML MoSe$_2$/31ML FePS$_3$ and individual MoSe$_2$ monolayer at 15 K. The peaks at 1.66 and 1.63 eV in both spectra correspond to neutral exciton (X) and charged trion (T) emission from the MoSe$_2$ monolayer, respectively. The appearance of the trion emission in the individual MoSe$_2$ monolayer is due to the unintentional *n*-type doping as a consequence of impurities attached to reactive chalcogenide vacancies or defects in the SiO$_2$ substrate.[44, 45, 46] In Figure 3 (d), we observe a giant enhancement of the trion emission and a dramatic suppression of the exciton emission (the PL intensity ratio $\frac{I_{Trion}}{I_X} = \sim 20$) for MoSe$_2$ in the heterojunction in comparison with $\frac{I_{Trion}}{I_X} = \sim 0.44$ for individual 1ML MoSe$_2$. The strong enhancement of the trion emission implies a strong doping of MoSe$_2$ in vdWH. Again, this agrees well with the definition of type-I band alignment where carriers can flow only in one direction from WBG to NBG. In heavily doped semiconductors excitons can trap free electrons or holes to form trions. This phenomenon also can be explained by a model based on the law of mass action:[47, 48]

$$\frac{N_X n_e}{N_T} = \left(\frac{4 m_X m_e}{\pi \hbar^2 m_T}\right) k_B T \, exp(-\frac{E_B}{k_B T})$$

Here $N_T$, $N_X$ and $n_e$ are trion, exciton and charge carrier concentrations, $\hbar$ is the reduced Planck's constant, $k_B$ is the Boltzmann constant, $T$ is the temperature, $E_B$ is the trion binding energy, $m_e$ is effective mass of electrons, $m_X$ and $m_T$ are exciton and trion effective masses, respectively. According to this model, there is a positive correlation between $\frac{N_T}{N_X}$ and $n_e$, i.e. $\frac{N_T}{N_X}$ increases with increasing carrier concentration $n_e$. Since the concentrations of trions and excitons are proportional to their PL intensities ($\frac{N_T}{N_X} \propto \frac{I_T}{I_X}$), it means that doping (higher $n_e$ or $n_p$) can lead to a higher PL intensity ratio $\frac{I_T}{I_X}$. Besides, the increased trion binding energy, which can be modified by the dielectric environment, could also lead to a higher $\frac{I_T}{I_X}$. Specifically, a lower dielectric constant of the environment can reduce the total screening effect on 1ML TMDCs, which can result in a higher trion binding energy and further can contribute to a higher $\frac{I_T}{I_X}$.[35] Due to $\varepsilon_{FePS_3} < \varepsilon_{SiO_2}$, it means that FePS$_3$ in vdWHs may provide an environment that is more energetically favorable for trion formation in MoSe$_2$/FePS$_3$ than in individual MoSe$_2$ being directly contacted with a SiO$_2$ substrate.

Figure 3 (e) shows the normalized TDPL spectra for vdWHs (blue-shaded areas) and individual 1ML MoSe$_2$ (black-shaded areas). The red shift of exciton and trion peaks for both samples is observed with increasing temperature, which is due to temperature-induced bandgap shrinkage. Below 50 K, the trion emission gives a pronounced peak and dominates the spectra of vdWH, while in individual 1ML MoSe$_2$ the exciton emission dominates the PL spectra because of the lower carrier concentration and thus lower trion concentration. At 50 K the trion and exciton emissions from the heterojunction have comparable intensity, but with increasing temperature the trion emission vanishes. This is due to thermal fluctuations resulting in the conversion from trions to excitons. In individual 1ML MoSe$_2$ the trion emission is already negligible as compared with the exciton peak at 50 K as a consequence of lower carrier concentration. In the temperature range of 90-300 K, only the exciton emission peak is observed for both vdWHs and individual MoSe$_2$. Note that we assume FePS$_3$ has no effect on the exciton peak position of MoSe$_2$ in vdWHs. In addition, the thermal stability of the trion also depends on its charge state. Positively charged trions, X$^+$, have localized holes that are less prone to scattering, leading to the reduced line width.[26]

In order to clarify the charge sign of trions in both systems and to explore their potential impacts on valleytronics, we have employed PL excitation with circularly polarized light (see **Figure 4**). Using left-($\sigma^-$) or right-handed ($\sigma^+$) excitation in both systems, we have seen differences in the PL emission. Here, the degree of circular polarization is evaluated as $P_c = [I(\sigma+) - I(\sigma-)]/[I(\sigma+) + I(\sigma-)]$, where $I(\sigma+)$ and $I(\sigma-)$ correspond to the PL intensities under left-($\sigma^-$) or right-handed ($\sigma^+$) excitation. In the individual monolayer MoSe$_2$ the exciton emission is enhanced by 13% for $\sigma^+$ excitation while the trion emission is almost not affected with a small broadening. The exciton emission polarization is resulted from the broken inversion symmetry and the strong SOC in monolayer TMDCs,[49] which gives rise to two inequivalent K+ and K- valleys with different population. Due to the spin-valley locking effect, the right-handed ($\sigma^+$) circular polarized photon initializes a carrier in the K+ valley and the left-handed ($\sigma^-$) initializes a carrier in the K- valley, i.e. valley dependent optical selection rules.[50, 51] Zeng et.al. reported that 30% valley polarization can be achieved in pristine monolayer MoS$_2$ with circularly polarized light pump.[52] For the PbI$_2$/WS$_2$ heterostructure with type-I band alignment, Zhang and coworkers have shown near unity spin polarization in PbI$_2$ due to the modification of the carrier lifetime.[51] In the heterostructure MoSe$_2$/FePS$_3$, the exciton emission is the same for both excitations, while the trion emission exhibits a polarization of about 14%. It suggests that no carrier spin polarization is introduced from Ising-type antiferromagnetic $FePS_3$ to $MoSe_2$. Moreover the trion emission from the heterostructure is narrower than from the individual $MoSe_2$ monolayer and the peak position is blue shifted for the heterostructure by about 2 meV. Since holes are more localized than electrons in $MoSe_2$, $X^+$ are immobile and less prone to scattering, leading to a narrow linewidth of $X^+$ than $X^-$.[26] A similar energy difference for the $X^+$ and $X^-$ emission and narrowing of the PL emission for $X^+$ in $MoSe_2$ were observed by Shepard et al.[26] Therefore, we can conclude that in our case negatively charged trions $X^-$ form at low temperature in the individual $MoSe_2$ monolayer, while in the heterostructures the trions are positively charged due to hole injection from *p*-type $FePS_3$ into the $MoSe_2$ monolayer.

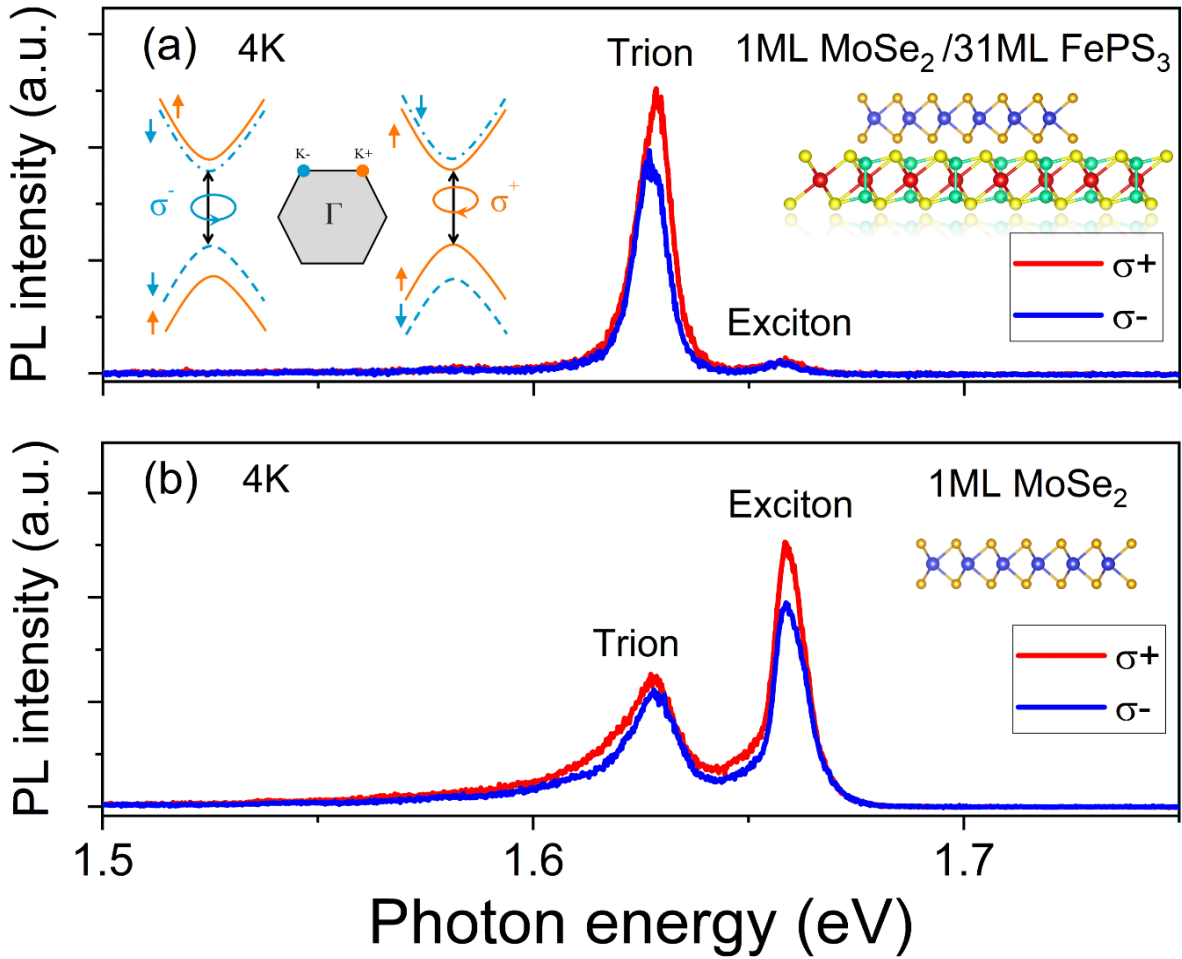

**Fig. 4.** Low temperature PL emission after excitation with left- or right-handed circularly polarized light obtained from the heterojunction **(a)** and from the individual $MoSe_2$ monolayer **(b)**. The inset shows the chiral optical selection rules for interband optical transitions in nonequivalent valleys K+ and K- in monolayer $MoSe_2$.

## 3. Conclusion

In conclusion, we have performed a systematic study of optical properties for a novel semiconductor-antiferromagnetic $MoSe_2$/$FePS_3$ vdWH. DFT calculations show the type-I band alignment in ultrathin $MoSe_2$/$FePS_3$, which is verified by the PL results. The room temperature PL shows 20-fold enhancement of the emission from 1ML $MoSe_2$ in $MoSe_2$/$FePS_3$ vdWH, which is due to carrier injection from WBG $FePS_3$ to NBG $MoSe_2$. At low temperatures, the intensity ratio between trion and exciton emission increases from 0.44 to about 20. Moreover, the heterojunction exhibits trion emission polarization, which makes it attractive for valleytronics. Our results present a novel approach to modify the population of trions and excitons as well as the trion charge state by forming a heterostructure with type-I band alignment. This approach is transferable to many other 2D materials and is opening a broad playground for fundamental physics and for valleytronic applications.

## 4. Methods

*Sample preparation:* The 2D material flakes were mechanically exfoliated through blue tape and then transferred onto $SiO_2$/Si substrates (90 nm thick $SiO_2$ on heavily doped Si) by using a polymer stamp transfer technique within a nitrogen atmosphere in a glove box environment. This technique provides good crystalline quality, intact atomic interfaces and less contamination of the flake in comparison with other transfer techniques.[53] Also, pre-patterned electrodes with Ti/Au (5/45nm) on $SiO_2$/Si substrates were fabricated by standard photolithography and electron-beam evaporation processes. To fabricate the $MoSe_2$/$FePS_3$ vdWH, $FePS_3$ was first transferred onto the $SiO_2$/Si substrate, and then the monolayer $MoSe_2$ was transferred onto the top of the $FePS_3$ flake. For reference, the single $MoSe_2$ monolayer flakes were prepared in the same way. Finally, both the vdWH and single $MoSe_2$ flakes were covered with h-BN in order to reduce flake contamination and degradation resulting from contact with air.

*Characterization*: For the micro-PL and micro-Raman measurements, we used a cw frequency-doubled Nd:YAG laser for excitation at a wavelength of $\lambda$=532 nm. The laser power was controlled by introducing neutral density filters. The maximum laser power used here was 3.2 mW to avoid heating of the flake. The spot diameter of the laser on the sample was approximately 3 μm, i.e., smaller than the size of the target flakes. Both PL and micro-Raman investigations were performed in the temperature range from 4 to 300 K using a liquid-He cooled chamber and temperature controller. A liquid nitrogen-cooled Si-CCD camera was used to detect the PL emission, which is dispersed in a spectrometer. To check the PL polarization of trion and exciton in MoSe$_2$, the excitation with circular left-(right-) handed polarized light was performed. The measurement was done at 4 K. The thickness of FePS$_3$ is measured with atomic force microscopy (*AFM*) with a Bruker Multimode 8 system, and the thicknesses of MoSe$_2$ were estimated by optical contrast and confirmed by micro-Raman, PL results.

*Calculation details*: Spin-polarized density functional theory (DFT) calculations were performed in the framework of the projector augmented wave method using the VASP code.[54] The generalized gradient approximation (GGA) with the Perdew-Burke-Ernzerhof (PBE) parametrization was used for the exchange-correlation functional.[55] To account for the strong correlation effects among d-orbital electrons of iron atoms, the DFT + U method with an effective Hubbard value (U) of 4.5 eV for Fe atoms was employed. The plane-wave basis set with a kinetic energy cutoff of 600 eV was used. The Brillouin zone of the system was sampled using $8 \times 8 \times 1$ *k*-mesh for primitive cells and $5 \times 5 \times 1$ *k*-mesh for supercells. London dispersion interactions were included in the total energy as proposed by Grimme in the DFT-D2 method.[56] The geometry optimization was carried out until the atomic forces were less than 0.01 eV/Å. A vacuum layer of 20 Å perpendicular to the basal plane was introduced to implement the isolated slab condition. In order to adopt zigzag antiferromagnetic ordering in FePS$_3$, a unit cell of tetragonal shape with 20 atoms was considered, as shown in Figure 2(a). The heterostructures were constructed by using an interface consisting of $1 \times 1$ unit cell of FePS$_3$ and $1 \times 3$ unit cells of monolayer MoSe$_2$ corresponding to a lattice mismatch of only 0.6%. The supercells contain 38 atoms in 2 layers and 74 atoms in 4 layers (MoSe$_2$ triple layers+ FePS$_3$ monolayer) configurations.

**Acknowledgements**
The author J.M. Duan thanks China Scholarship Council (File No. 201706890037). L. Hu thanks the National Natural Science Foundation of China (project number 61804098) and the Zhejiang Provincial Natural Science Foundation of China (project number LZ21E020002). Y.J. Zeng thanks the Shenzhen Science and Technology Project under Grant Nos. JCYJ20180507182246321. A.V. Krasheninnikov also thanks the DFG for support within the projects KR 4866/2-1 (project number 339 406129719). The computational support from the Technical University of Dresden computing cluster (TAURUS) and from High Performance Computing Center (HLRS) in Stuttgart, Germany are gratefully appreciated. We would like to thank Mr Scheumann for the metal deposition of the substrates. The nanofabrication facilities (NanoFaRo) at the Ion Beam Center at the HZDR are also gratefully acknowledged.

## References

[1] A. Splendiani, L. Sun, Y. Zhang, T. Li, J. Kim, C.-Y. Chim, G. Galli, F. Wang, *Nano Lett.* **2010**, *10*, 1271.

[2] K. F. Mak, C. Lee, J. Hone, J. Shan, T. F. Heinz, *Phys. Rev. Lett.* **2010**, *105*, 136805.

[3] W. Zhao, Z. Ghorannevis, L. Chu, M. Toh, C. Kloc, P.-H. Tan, G. Eda, *ACS Nano* **2013**, *7*, 791.

[4] K. F. Mak, K. He, C. Lee, G. H. Lee, J. Hone, T. F. Heinz, J. Shan, *Nat. Mater.* **2013**, *12*, 207.

[5] A. M. Jones, H. Yu, N. J. Ghimire, S. Wu, G. Aivazian, J. S. Ross, B. Zhao, J. Yan, D. G. Mandrus, D. Xiao, *Nat. Nanotechnol.* **2013**, *8*, 634.

[6] J. S. Ross, S. Wu, H. Yu, N. J. Ghimire, A. M. Jones, G. Aivazian, J. Yan, D. G. Mandrus, D. Xiao, W. Yao, *Nat. Commun.* **2013**, *4*, 1.

[7] A. Singh, G. Moody, K. Tran, M. E. Scott, V. Overbeck, G. Berghäuser, J. Schaibley, E. J. Seifert, D. Pleskot, N. M. Gabor, *Phys. Rev. B* **2016**, *93*, 041401.

[8] K. F. Mak, K. He, J. Shan, T. F. Heinz, *Nat. Nanotechnol.* **2012**, *7*, 494.

[9] Y. Liu, N. O. Weiss, X. Duan, H.-C. Cheng, Y. Huang, X. Duan, *Nat. Rev. Mater.* **2016**, *1*, 1.

[10] F. Xia, H. Wang, D. Xiao, M. Dubey, A. Ramasubramaniam, *Nat. Photonics* **2014**, *8*, 899.

[11] K. Novoselov, o. A. Mishchenko, o. A. Carvalho, A. C. Neto, *Science* **2016**, *353*, 9439.

[12] F. Ceballos, M. Z. Bellus, H.-Y. Chiu, H. Zhao, *ACS Nano* **2014**, *8*, 12717.

[13] X. Hong, J. Kim, S.-F. Shi, Y. Zhang, C. Jin, Y. Sun, S. Tongay, J. Wu, Y. Zhang, F. Wang, *Nat. Nanotechnol.* **2014**, *9*, 682.

[14] Y. Yu, S. Hu, L. Su, L. Huang, Y. Liu, Z. Jin, A. A. Purezky, D. B. Geohegan, K. W. Kim, Y. Zhang, *Nano Lett.* **2015**, *15*, 486.

[15] M.-H. Chiu, M.-Y. Li, W. Zhang, W.-T. Hsu, W.-H. Chang, M. Terrones, H. Terrones, L.-J. Li, *ACS Nano* **2014**, *8*, 9649.

[16]  T. Roy, M. Tosun, X. Cao, H. Fang, D.-H. Lien, P. Zhao, Y.-Z. Chen, Y.-L. Chueh, J. Guo, A. Javey, *ACS Nano* **2015**, *9*, 2071.

[17]  P. Rivera, J. R. Schaibley, A. M. Jones, J. S. Ross, S. Wu, G. Aivazian, P. Klement, K. Seyler, G. Clark, N. J. Ghimire, *Nat. Commun.* **2015**, *6*, 1.

[18]  F. Ceballos, M. Z. Bellus, H.-Y. Chiu, H. Zhao, *Nanoscale* **2015**, *7*, 17523.

[19]  W.-T. Hsu, Z.-A. Zhao, L.-J. Li, C.-H. Chen, M.-H. Chiu, P.-S. Chang, Y.-C. Chou, W.-H. Chang, *ACS Nano* **2014**, *8*, 2951.

[20]  M. Z. Bellus, M. Li, S. D. Lane, F. Ceballos, Q. Cui, X. C. Zeng, H. Zhao, *Nanoscale Horiz.* **2017**, *2*, 31.

[21]  T. Yamaoka, H. E. Lim, S. Koirala, X. Wang, K. Shinokita, M. Maruyama, S. Okada, Y. Miyauchi, K. Matsuda, *Adv. Funct. Mater.* **2018**, *28*, 1801021.

[22]  H. Cai, J. Kang, H. Sahin, B. Chen, A. Suslu, K. Wu, F. Peeters, X. Meng, S. Tongay, *Nanotechnology* **2016**, *27*, 065203.

[23]  W. Zheng, B. Zheng, Y. Jiang, C. Yan, S. Chen, Y. Liu, X. Sun, C. Zhu, Z. Qi, T. Yang, *Nano Lett.* **2019**, *19*, 7217.

[24]  Y. H. Zhou, Z. B. Zhang, P. Xu, H. Zhang, B. Wang, *Nanoscale Res. Lett.* **2019**, *14*, 1.

[25]  J. J. Carmiggelt, M. Borst, T. van der Sar, *Sci. Rep.* **2020**, *10*, 1.

[26]  G. D. Shepard, J. V. Ardelean, O. A. Ajayi, D. Rhodes, X. Zhu, J. C. Hone, S. Strauf, *ACS Nano* **2017**, *11*, 11550.

[27]  P. Tonndorf, R. Schmidt, P. Böttger, X. Zhang, J. Börner, A. Liebig, M. Albrecht, C. Kloc, O. Gordan, D. R. Zahn, *Opt. Express* **2013**, *21*, 4908.

[28]  D. Nam, J.-U. Lee, H. Cheong, *Sci. Rep.* **2015**, *5*, 17113.

[29]  J.-U. Lee, S. Lee, J. H. Ryoo, S. Kang, T. Y. Kim, P. Kim, C.-H. Park, J.-G. Park, H. Cheong, *Nano Lett* **2016**, *16*, 7433.

[30]  X. Wang, K. Du, Y. Y. F. Liu, P. Hu, J. Zhang, Q. Zhang, M. H. S. Owen, X. Lu, C. K. Gan, P. Sengupta, *2D Mater.* **2016**, *3*, 031009.

[31]  T. Björkman, A. Gulans, A. V. Krasheninnikov, R. M. Nieminen, *Phys. Rev. Lett.* **2012**, *108*, 235502.

[32]  Y. Gao, S. Lei, T. Kang, L. Fei, C.-L. Mak, J. Yuan, M. Zhang, S. Li, Q. Bao, Z. Zeng, *Nanotechnology* **2018**, *29*, 244001.

[33]  X. Zhang, X. Zhao, D. Wu, Y. Jing, Z. Zhou, *Adv. Sci.* **2016**, *3*, 1600062.

[34]  F. Bussolotti, H. Kawai, Z. E. Ooi, V. Chellappan, D. Thian, A. L. C. Pang, K. E. J. Goh, *Nano Futures* **2018**, *2*, 032001.

[35]  J. Wierzbowski, J. Klein, F. Sigger, C. Straubinger, M. Kremser, T. Taniguchi, K. Watanabe, U. Wurstbauer, A. W. Holleitner, M. Kaniber, *Sci. Rep.* **2017**, *7*, 1.

[36]  B. Zheng, W. Zheng, Y. Jiang, S. Chen, D. Li, C. Ma, X. Wang, W. Huang, X. Zhang, H. Liu, *J. Am. Chem. Soc.* **2019**, *141*, 11754.

[37]  Y. Gao, S. Lei, T. Kang, L. Fei, C.-L. Mak, J. Yuan, M. Zhang, S. Li, Q. Bao, Z. Zeng, *Nanotechnology* **2018**, *29*, 244001.

[38]  X. Wang, Y. Gong, G. Shi, W. L. Chow, K. Keyshar, G. Ye, R. Vajtai, J. Lou, Z. Liu, E. Ringe, *ACS Nano* **2014**, *8*, 5125.

[39]  D. Sercombe, S. Schwarz, O. Del Pozo-Zamudio, F. Liu, B. Robinson, E. Chekhovich, I. Tartakovskii, O. Kolosov, A. Tartakovskii, *Sci. Rep.* **2013**, *3*, 3489.

[40]  W. Zheng, B. Zheng, C. Yan, Y. Liu, X. Sun, Z. Qi, T. Yang, Y. Jiang, W. Huang, P. Fan, *Adv. Sci.* **2019**, *6*, 1802204.

[41]  X. Zong, H. Hu, G. Ouyang, J. Wang, R. Shi, L. Zhang, Q. Zeng, C. Zhu, S. Chen, C. Cheng, *Light Sci. Appl.* **2020**, *9*, 1.

[42]  R. Cheng, D. Li, H. Zhou, C. Wang, A. Yin, S. Jiang, Y. Liu, Y. Chen, Y. Huang, X. Duan, *Nano Lett* **2014**, *14*, 5590.

[43]  M. Van der Donck, F. Peeters, *Phys. Rev. B* **2018**, *98*, 115104.

[44]  K. Dolui, I. Rungger, S. Sanvito, *Phy. Rev. B* **2013**, *87*, 165402.

[45]  A. Singh, A. K. Singh, *Phys. Rev. B* **2019**, *99*, 121201.

[46]  H.-P. Komsa, J. Kotakoski, S. Kurasch, O. Lehtinen, U. Kaiser, A. V. Krasheninnikov, *Phys. Rev. Lett.* **2012**, *109*, 035503.

[47]  J. Siviniant, D. Scalbert, A. Kavokin, D. Coquillat, J. Lascaray, *Phys. Rev. B* **1999**, *59*, 1602.

[48]  N. Peimyoo, W. Yang, J. Shang, X. Shen, Y. Wang, T. Yu, *ACS Nano* **2014**, *8*, 11320.

[49]  R. Suzuki, M. Sakano, Y. Zhang, R. Akashi, D. Morikawa, A. Harasawa, K. Yaji, K. Kuroda, K. Miyamoto, T. Okuda, *Nat. Nanotechnol.* **2014**, *9*, 611.

[50]  T. Cao, G. Wang, W. Han, H. Ye, C. Zhu, J. Shi, Q. Niu, P. Tan, E. Wang, B. Liu, *Nat. Commun.* **2012**, *3*, 887.

[51]  H. Zeng, J. Dai, W. Yao, D. Xiao, X. Cui, *Nat. Nanotechnol.* **2012**, *7*, 490.

[52]  D. Zhang, Y. Liu, M. He, A. Zhang, S. Chen, Q. Tong, L. Huang, Z. Zhou, W. Zheng, M. Chen, *Nat. Commun.* **2020**, *11*, 1.

[53]  M. Velický, P. S. Toth, *Appl. Mater. Today* **2017**, *8*, 68.

[54]  G. Kresse, J. Hafner, *Phys. Rev. B* **1993**, *47*, 558.

[55] J. P. Perdew, K. Burke, M. Ernzerhof, *Phys. Rev. Lett.* **1996**, *77*, 3865.

[56] S. Grimme, *J. Comput. Chem.* **2006**, *27*, 1787.

# Supporting Information

**Enhanced trion emission in monolayer MoSe$_2$ by constructing a type-I van der Waals heterostructure**

Juanmei Duan[*], Phanish Chava, Mahdi Ghorbani-Asl, Arkady V. Krasheninnikov, Denise Erb, Liang Hu, Harald Schneider, Lars Rebohle, Artur Erbe, Manfred Helm, Yu-Jia Zeng, Shengqiang Zhou[*] and Slawomir Prucnal[*]

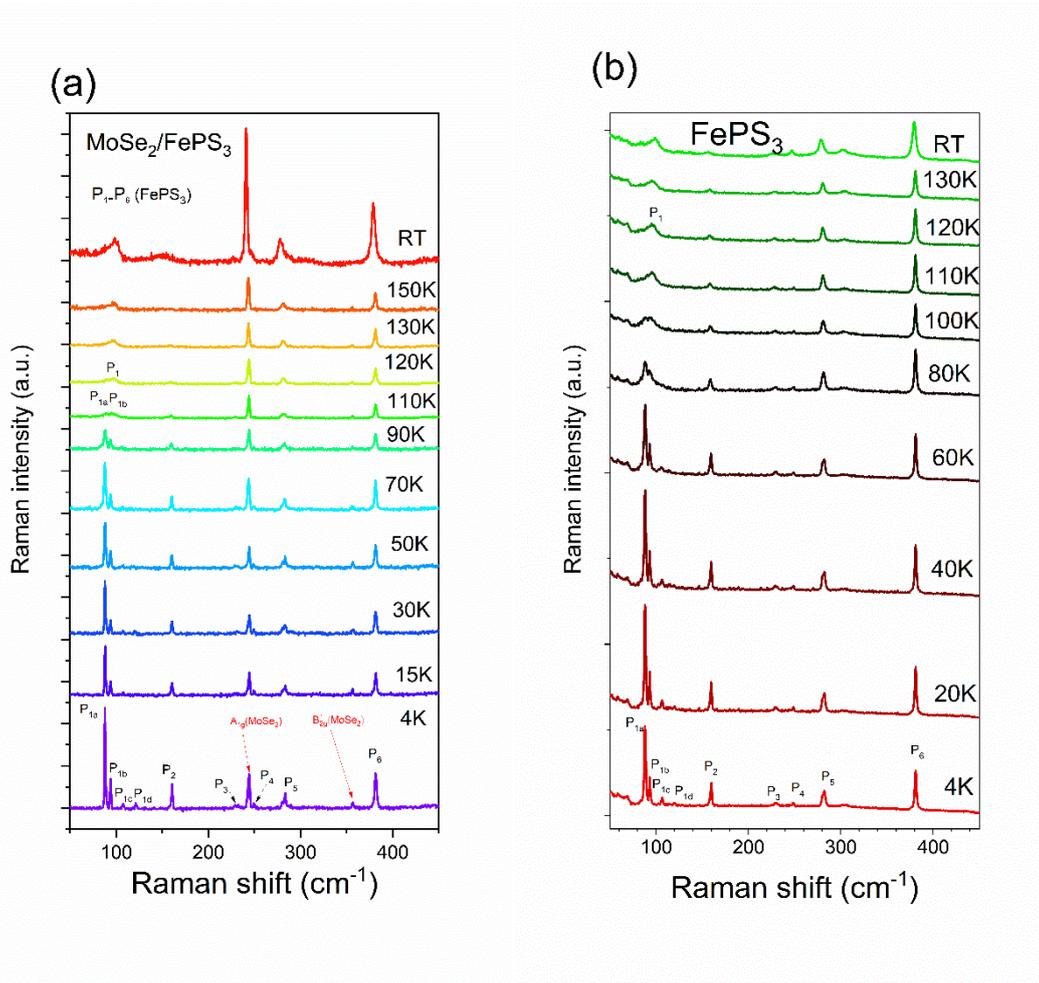

Figure S1. (a) Temperature dependent Raman spectra of a multilayer MoSe$_2$/FePS$_3$ heterostructure and (b) Raman spectra obtained from an individual FePS$_3$ flake.

Figure S1 shows the temperature-dependent Raman spectra from multilayer MoSe$_2$/FePS$_3$ heterostructure and multilayer FePS$_3$. A$_{1g}$ and $B_{2g}^1$ peaks of MoSe$_2$ shift to lower wavenumber with increasing temperature as shown in Figure S1 (a). The peaks labeled with P$_1$-P$_6$ originate from FePS$_3$ in the heterojunction. Specifically, the P$_3$-P$_6$ phonon modes are assigned to the (P$_2$S$_6$)$^{4-}$ bipyramid structures and P$_1$-P$_2$ are related to the Fe atoms and are sensitive to the magnetic coupling between Fe atoms. Note that at low temperature the $P_1$ phonon mode shows several separated sharp peaks including $P_{1a}$-$P_{1d}$, which become broad and asymmetric with increasing temperature. Especially as temperature increases from 110 to 120 K, $P_{1a}$ and $P_{1b}$ merge to one broad peak $P_1$. This phenomenon is due to the transition of the magnetic state from antiferromagnetic (AF) to paramagnetic (PM) in FePS$_3$ with a Néel temperature at 118 K.[1] Lee *et al* reported that the magnetic state exists down to the FePS$_3$ monolayer limit using temperature-dependent Raman measurement.[2] Figure S1 (b) illustrates the Raman spectra of multilayer FePS$_3$. It shows a similar behavior for $P_1$ mode like FePS$_3$ in MoSe$_2$/FePS$_3$ with increasing temperature to 120 K, which suggests the Néel temperature remains at around 118 K for single multilayer FePS$_3$ or after combined with MoSe$_2$.

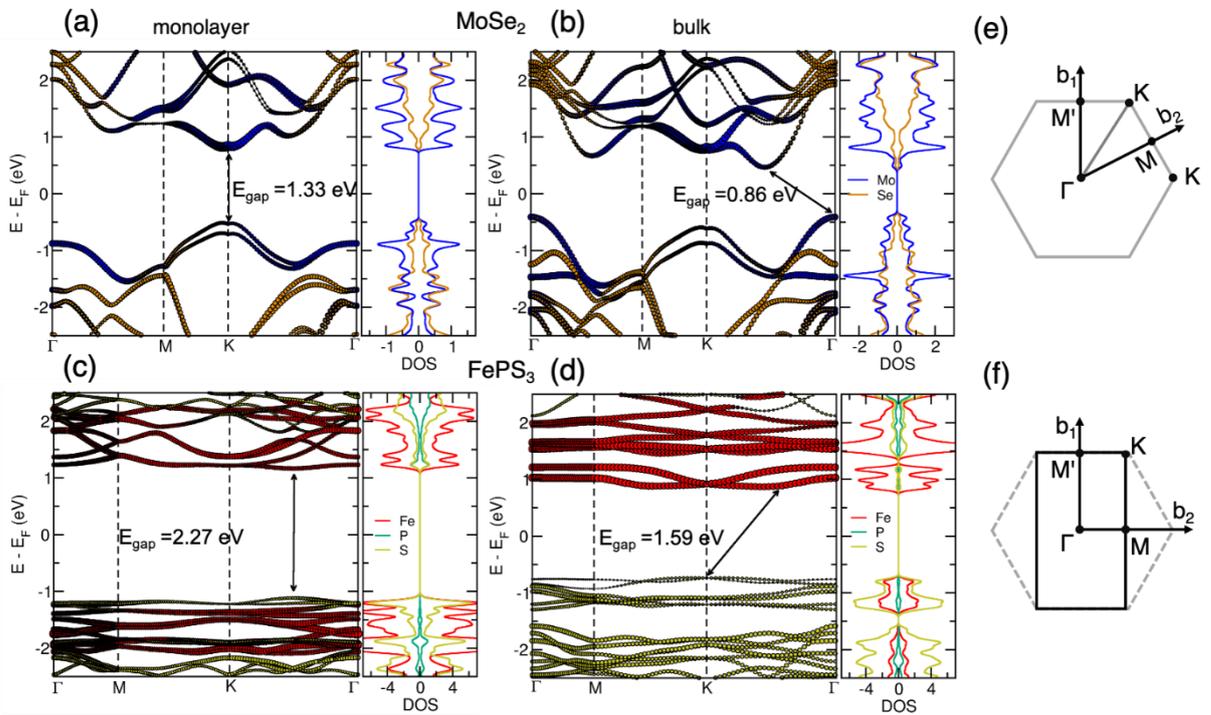

Figure S2. (a,b) Electronic band structure and density of states (DOS) of 1ML MoSe$_2$ and bulk MoSe$_2$ calculated at the DFT/PBE level. The spin-orbit coupling (SOC) is considered. (c,d) Electronic band structure and density of states (DOS) of 1ML FePS$_3$ and bulk FePS$_3$ calculated using the DFT+U method. (e,f) First Brillouin zone and high-symmetry k-points of hexagonal and rectangular unit cells.

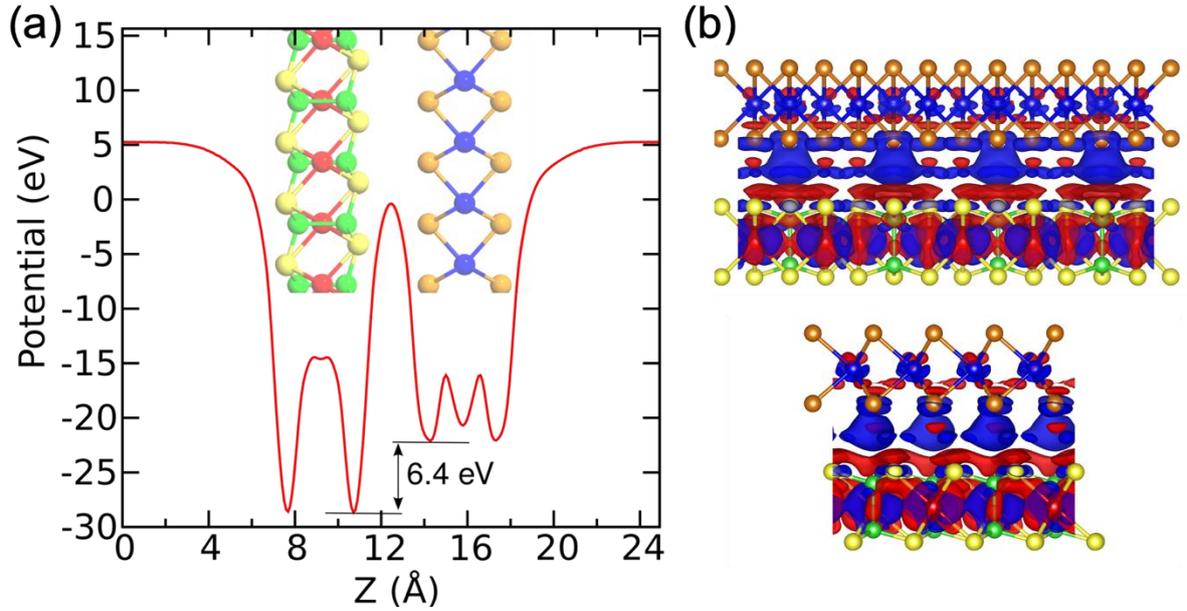

Figure S3. (a) Averaged electronic potential along the perpendicular direction of the FePS$_3$/MoSe$_2$ heterostructure. (b) Difference of the electron densities in the heterostructure and isolated sheets as viewed from two directions. Blue areas correspond to an increase in electron density, red to the depletion of the electronic charge.

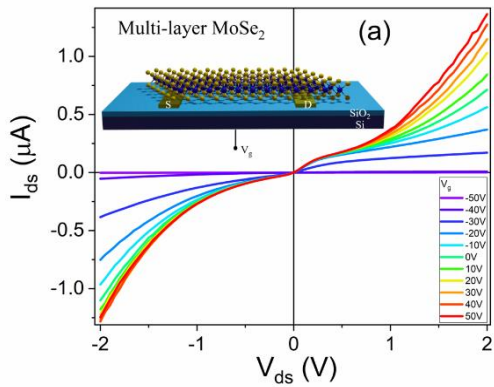
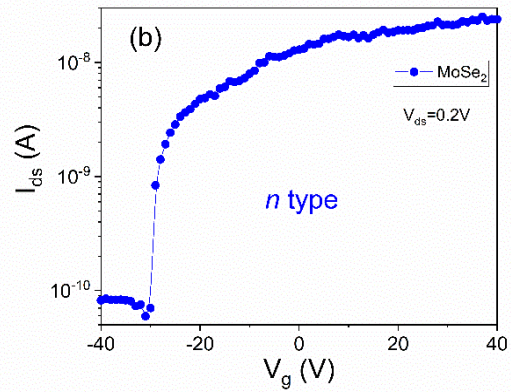
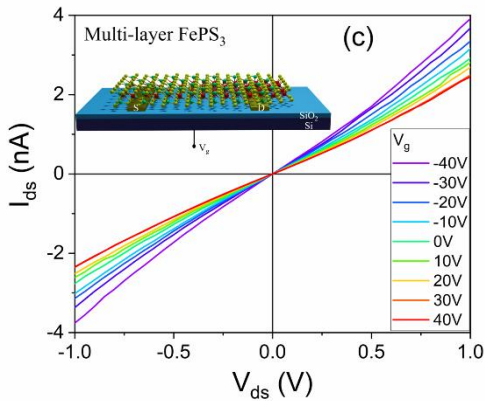
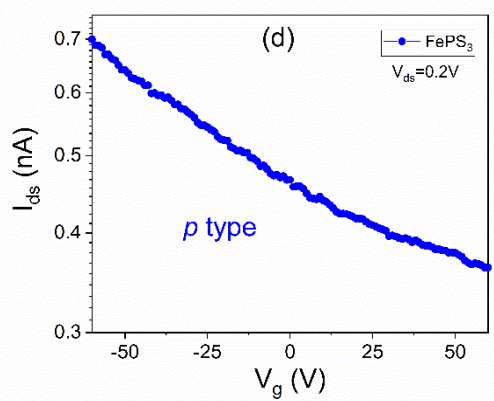

Figure S4. (a) Output characteristics of a multilayer MoSe$_2$ transistor at different gate voltages. b) Transfer characteristics of a multilayer MoSe$_2$ transistor measured with V$_{ds}$ = 0.2 V. (c) Output characteristics of a multilayer FePS$_3$ transistor at different gate voltages. (d) Transfer characteristics of a multilayer FePS$_3$ transistor measured with V$_{ds}$ = 0.2 V. The insets in Figure S4 (a) and Figure S4 (c) represents the schematic of MoSe$_2$ and FePS$_3$ transistors.

Figure S4 (a) shows that the I$_{ds}$ increases with increasing positive gate voltage suggesting *n*-type conductivity of MoSe$_2$. Figure S4 (b) presents the transfer characteristics of multilayer MoSe$_2$ and the on-off ratio of MoSe$_2$ is estimated to be ~10$^2$ from the curve. Moreover, we calculated the field-effect mobility of MoSe$_2$ using the equation $\mu_{FE} = \left(\frac{L}{WC_{ox}V_{ds}}\right)(d\,I_{ds}/d\,V_g)$, where $C_{ox}$ =3.84×10$^{-8}$ F/cm$^2$ is the capacitance per unit area of 90 nm SiO$_2$ between MoSe$_2$ and the back gate, L and W are the channel length of 23 μm and channel width of 20 μm. The calculated field-effect mobility of MoSe$_2$ is determined to be $\mu_{FE}$ = 3.1×10$^{-2}$ cm$^2$ V$^{-1}$ S$^{-1}$ at V$_{ds}$ = 0.2 V. It is worth mentioning that $\mu_{FE}$ of MoSe$_2$ is dependent on applied V$_{ds}$ and higher V$_{ds}$ can lead to a higher $\mu_{FE}$. Figure S4 (c) and (d) demonstrates the output characteristics and transfer output characteristics of a multilayer FePS$_3$ transistor at different gate voltages. I$_{ds}$ of FePS$_3$ increases with increasing negative gate voltage suggesting *p*-type conductivity of FePS$_3$.

# References

[1]    X. Wang, K. Du, Y. Y. F. Liu, P. Hu, J. Zhang, Q. Zhang, M. H. S. Owen, X. Lu, C. K. Gan, P. Sengupta, *2D Mater.* **2016**, *3*, 031009.

[2]    J.-U. Lee, S. Lee, J. H. Ryoo, S. Kang, T. Y. Kim, P. Kim, C.-H. Park, J.-G. Park, H. Cheong, *Nano Lett* **2016**, *16*, 7433.